\documentclass[preprintnumbers,amsmath,amssymb,floatfix,11pt,prd,onecolumn,
superscriptaddress,nofootinbib]{revtex4}
\usepackage{latexsym}
\usepackage{epsfig}
\usepackage{epstopdf}
\usepackage{amssymb}
\usepackage{amsmath}
\usepackage{comment}
\usepackage{lineno,hyperref,pdflscape}
\usepackage{float}

\newcommand{\be}{\begin{equation}}
\newcommand{\ee}{\end{equation}}
\newcommand{\ben}{\begin{eqnarray}}
\newcommand{\een}{\end{eqnarray}}

\newcommand{\p}{\partial}

\begin{document}

\title{New exact solutions of Bianchi I, Bianchi III and Kantowski-Sachs spacetimes
in scalar-coupled gravity theories via Noether gauge symmetries}

\author{U. Camci}
\email{ucamci@akdeniz.edu.tr}\affiliation{Department of Physics, Akdeniz University, 07058 Antalya, Turkey}

\author{A. Yildirim}
\email{aydinyildirim@akdeniz.edu.tr}\affiliation{Department of Physics, Akdeniz University, 07058 Antalya, Turkey}

\author{I. Basaran Oz}
\email{isilbasaran@akdeniz.edu.tr}\affiliation{Department of Physics, Akdeniz University, 07058 Antalya, Turkey}

\begin{abstract}
The Noether symmetry approach is useful tool to restrict the arbitrariness in a gravity theory when the equations of motion are underdetermined due to the high number of functions to be determined in the {\it ansatz}. We consider two scalar-coupled theories of gravity, one motivated by {\it induced gravity}, the other more standard; in Bianchi I, Bianchi III and Kantowski-Sachs cosmological models. For these models, we present a full set of Noether gauge symmetries, which are more general than those obtained by the strict Noether symmetry approach in our recent work. Some exact solutions are derived using the first integrals corresponding to the obtained Noether gauge symmetries.
\end{abstract}

\maketitle



\section{Introduction}

Cosmological problems such as dark matter, dark energy, the flatness and horizon problems, deriving from the combination of general relativity (GR) and the standard model of particle physics, led to the realization that the ``standard'' model of cosmology is not sufficient to describe the universe in extreme regimes \cite{capo2014}. Alternative theories of gravity, which must agree with GR in the weak field regime \cite{fujii}, might be able to solve these problems. Scalar-tensor theories of gravity, including scalar fields non-minimally coupled to gravity, have created considerable interest in cosmology because they introduce naturally scalar fields which are capable of giving rise to inflationary behaviour \cite{starobinsky,guth} of the universe, and generate dark energy dynamics \cite{capo2008b}.

Another motivation for considering scalar-tensor theories is the idea of {\it emergent} or {\it induced gravity}, originally proposed by Sakharov and Zeldovich \cite{sakharov,zeldovich}, and refined by
Adler and Zee \cite{adler-1,adler-2,adler-3,zee}. A particular realization of scalar-tensor theories, the $\Phi^{2}R$ theory, stands out in these discussions, and some of its consequences are worked out in \cite{pollock-1,pollock-2,sanyal02}. We will also work with this theory in the present study.

All cosmological models but the simplest Friedmann-Lemaitre-Robertson-Walker (FLRW) model contain more arbitrary functions than can be determined using the field equations for the matter and the geometry. Alternative theories of gravity may bring further unspecified functions as parts of the Lagrangian, increasing the arbitrariness. One suitable way to restrict this arbitrariness is to use
Noether symmetries which are directly related to the presence of conserved quantities, as selection rules. They can enable us to choose the scalar field's self-interaction potential in a dynamical way, and might be able to reduce the number of dynamical variables of the system of differential equations due to possible cyclic variables.

A Lagrangian density $\mathcal{L}$ admits a strict Noether symmetry if there exists a vector field ${\bf X}$, for which the Lie derivative of the Lagrangian  vanishes, i.e. $\pounds_{\bf X} \mathcal{L} = 0$.
This approach in the context of spacetimes like FLRW and Bianchi type universe models has already been used with great success in the frameworks of $f(R), f(T)$ and specific models of scalar-tensor gravity theories, allowing the identification of dynamical conserved quantities and the derivation of new exact solutions
\cite{demianski92,capo93,capo94a,capo94c,capo00,capo2007,capo2008a,capo2009,sanyal02,camci,camci2}. In particular, for the Kantowski-Sachs (KS) spacetime, the strict Noether and dynamical symmetries have been discussed by Sanyal \cite{sanyal02}, and generalized by Camci and Kucukakca \cite{camci} to  include Bianchi I (BI) and Bianchi III (BIII) spacetimes.

The strict Noether symmetry approach can be generalized to include a gauge term, giving the {\it Noether gauge symmetry} (NGS) approach,
a more suitable method to seek for physically motivated solutions of field equations; for some applications, see
\cite{camci3,Jamil:2012fs,Jamil:2012zm,sharif2013,yusuf,Aslam:2012tj,Jamil:2011pv}

In the present paper, we apply the NGS to BI, BIII and KS spacetimes, generalizing our earlier work \cite{camci}. The paper is organized as follows.
In the next section, we present the Lagrangian, rederive for completeness, both the equations of motion in  scalar-coupled theory of gravity, and their specialized forms for  BI, BIII and KS spacetimes.  In section \ref{ngs}, we discuss the NGS approach to the Lagrangian for these spacetimes, and give solutions of the NGS equations for the $\Phi^{2}R$ theory motivated from induced gravity, and the often-used $(1-\zeta \Phi^{2})R$ theory.  In section \ref{fieldeqns}, we search for exact solutions of the field equations using the obtained NGS. Finally, in section \ref{conc}, we conclude with a brief summary and discussion of the obtained results.

\section{The equations of motion}
\label{general}

The general form of the Lagrangian density for the action $\mathcal{A}=\int{ \mathcal{L} dt }$ involving gravity non-minimally coupled
with a real scalar field $\Phi$ is given by \cite{capo93}
\begin{eqnarray}
\mathcal{L} & = & \int{d^3 x \sqrt{-g} \left[ F(\Phi) R +
\frac{\epsilon}{2} g^{ab} \Phi_{a} \Phi_{b} - U(\Phi) \right]}
\label{action}
\end{eqnarray}
where $R$ is the Ricci scalar, $F(\Phi)$ is the generic function
describing the coupling, $U(\Phi)$ is the potential for the scalar
field, $\Phi_a = \Phi_{,a}$ stand for the components of
the gradient of $\Phi$ and the signature of metric is given by the
parameter $\epsilon = +1$ and $-1$ for signatures $(+ - - -)$ and $(- +
+ +)$, respectively. We use Planck units. For $F(\Phi) =
-1/2$, it reduces to the Einstein-Hilbert action minimally coupled with a scalar field.
For $F(\Phi)= \Phi^2 /6$, the conformally coupled theory can be obtained.
For $F(\Phi)= F_0\, \Phi^2 /12, (F_0 \neq \epsilon )$ represents the theory motivated from induced gravity, and for $F(\Phi)=1- \zeta \Phi^2$, $\mathcal{L}$ is of the form of the more standard non-minimally coupled scalar field theory.

The variation of the action $\mathcal{A}$  with respect to $g_{ab}$ provides the field
equations
\begin{eqnarray}
& & F(\Phi) G_{ab} = - \frac{\epsilon}{2} T^{\Phi}_{ab} - g_{ab} \Box
F(\Phi) + F(\Phi)_{;ab} \label{feq}
\end{eqnarray}
where $\Box$ is the d'Alembert operator, \be G_{ab} = R_{ab} -
\frac{1}{2} R g_{ab}\ee is the Einstein tensor, and
\begin{eqnarray}
& & T^{\Phi}_{ab} = \Phi_{a} \Phi_{b} - \frac{1}{2} g_{ab} \Phi_{c}
\Phi^{c} + \epsilon g_{ab} U(\Phi) \label{emt}
\end{eqnarray}
is the energy-momentum tensor of the scalar field. The
variation with respect to $\Phi$ gives rise to the Klein-Gordon
equation governing the dynamics of the scalar field
\begin{eqnarray}
& & \epsilon \Box \Phi - R F'(\Phi) + U'(\Phi) = 0, \label{kg}
\end{eqnarray}
where the prime indicates the derivative with respect to $\Phi$. The Bianchi identity $G^{ab}_{\,\,\,\,;b} = 0$, which gives the conservation laws for the scalar field, also yields the Klein-Gordon equation (\ref{kg}) as a general result \cite{capo94c}.

As discussed in the Introduction, we will treat the BI, BIII and KS
spacetimes. The line element for these spacetimes can be written in the common form
\begin{equation} \label{metric}
ds^2 = \epsilon \left( dt^2 - A^2 dr^2 \right) -\epsilon B^2
\left( d\theta^2 + \Sigma^2 (q,\theta) d\Phi^2 \right),
\end{equation}
where $A$ and $B$  depend on $t$ only, and $\Sigma(q,\theta) = \theta, \sinh\theta, \sin\theta$ for $q=0, -1,+1$, respectively. The expression (\ref{metric}) represents BI for $q = 0$, BIII for $q =-1$ and KS  for $q=1$.

The Ricci scalar computed from the line element is
\begin{equation}
R = -2 \epsilon \left[ \frac{\ddot{A}}{A} + 2 \frac{\ddot{B}}{B}+
\frac{\dot{B}^2}{B^2} + 2\frac{\dot{A}\dot{B}}{A B} +
\frac{q}{A^2} \right],
\end{equation}
where the dot represents the derivation with respect to time. Then, the Lagrangian  of BI, BIII and KS spacetimes becomes
\begin{eqnarray}
& & \mathcal{L} = 2 \epsilon F A \dot{B}^2 + 4 \epsilon F B \dot{A}\dot{B} +
2 \epsilon F' B^2 \dot{A} \dot{\Phi} + 4 \epsilon F' A B \dot{B}
\dot{\Phi} \nonumber \\& &  \qquad - 2 \epsilon q F A  + A B^2 \left[
\frac{\dot{\Phi}^2}{2} - U(\Phi) \right]. \label{lag-b13ks}
\end{eqnarray}
The field equations (\ref{feq}) and Klein-Gordon equation
(\ref{kg}) for the metric (\ref{metric}) become
\begin{equation}
\frac{\dot{B}^2}{B^2} + 2\frac{\dot{A} \dot{B}}{A B} +
\frac{q}{B^2} + \frac{F'}{F} \left( \frac{\dot{A}}{A} + 2
\frac{\dot{B}}{B} \right) \dot{\Phi} + \frac{\epsilon}{2 F}\left[
\frac{\dot{\Phi}^2}{2} + U(\Phi) \right] =0,  \label{feq1}
\end{equation}
\begin{equation}
2 \frac{\ddot{B}}{B} + \frac{\dot{B}^2}{B^2}
+ \frac{q}{B^2} + \frac{F'}{F} \left[ \ddot{\Phi} + 2
\frac{\dot{B}}{B} \dot{\Phi} \right] + \left( \frac{F''}{F} -
\frac{\epsilon}{4 F} \right) \dot{\Phi}^2  + \frac{\epsilon}{2 F}
U(\Phi)= 0, \label{feq2}
\end{equation}
\begin{equation}
\frac{\ddot{A}}{A} + \frac{\ddot{B}}{B}+ \frac{\dot{A}
\dot{B}}{A B} + \frac{F'}{F} \left[ \ddot{\Phi} + \left(
\frac{\dot{A}}{A} + \frac{\dot{B}}{B}\right) \dot{\Phi} \right]  +
\left( \frac{F''}{F} - \frac{\epsilon}{4 F} \right) \dot{\Phi}^2 +
\frac{\epsilon U(\Phi) }{2 F}  = 0, \quad \label{feq3}
\end{equation}
\begin{equation}
\frac{\ddot{A}}{A}+ 2\frac{\ddot{B}}{B} +
\frac{\dot{B}^2}{B^2} + 2\frac{\dot{A}\dot{B}}{A B} +
\frac{q}{B^2} + \frac{\epsilon}{2 F'}\left[ \ddot{\Phi} + \left(
\frac{\dot{A}}{A} + 2\frac{\dot{B}}{B}\right) \dot{\Phi} +
U'(\Phi) \right] = 0, \label{feq4}
\end{equation}
where $F' \neq 0$. Note that the equations (\ref{feq2})-(\ref{feq4}) can also be obtained as the Euler-Lagrange equations using the Lagrangian (\ref{lag-b13ks}), whereas eq.(\ref{feq1}), the (0,0)-Einstein equation, can be obtained as the requirement for the \emph{energy function} $E_{\mathcal{L}}$ associated with the Lagrangian (\ref{lag-b13ks})
\begin{eqnarray}
E_{\mathcal{L}} &=& \frac{\partial \mathcal{L}}{\partial \dot{A}} \dot{A} +
\frac{\partial \mathcal{L}}{\partial \dot{B}} \dot{B} +
\frac{\partial \mathcal{L}}{\partial \dot{\Phi}} \dot{\Phi} - \mathcal{L} \nonumber \\
&=& \frac{\dot{B}^2}{B^2} + 2\frac{\dot{A} \dot{B}}{A B} +
\frac{q}{B^2} + \frac{F'}{F} \left( \frac{\dot{A}}{A} + 2
\frac{\dot{B}}{B} \right) \dot{\Phi} + \frac{\epsilon}{2 F}\left[
\frac{\dot{\Phi}^2}{2} + U(\Phi) \right] \label{e-b13ks}
\end{eqnarray}
to vanish.

If $\ddot{A}$ and $\ddot{B}$ can be eliminated form the Eqs. (\ref{feq2})-(\ref{feq4}) the continuity equation can be found as
\begin{equation}
2 (3 F'^2 - \epsilon F) \left[ \ddot{\Phi} + \frac{\dot{A}}{A} \dot{\Phi} + 2 \frac{\dot{B}}{B} \dot{\Phi} \right] + F' ( 6 F'' - \epsilon) \dot{\Phi} + 2 \epsilon ( 2 U F' - F U' ) = 0.  \label{cont}
\end{equation}

The unknown quantities of the field equations are $A, B, \Phi, U(\Phi)$ and $F(\Phi)$, but we have only four independent differential equations, namely, Eqs. (\ref{feq1})-(\ref{feq4}). Then, in order to solve this system of nonlinear differential equations we need to assume a functional form of the scalar field potential energy $U(\Phi)$ or the function $F(\Phi)$.

\section{Noether Gauge Symmetries} \label{ngs}

For the KS metrics of the signature $(+2)$, the forms of coupling of the scalar field and the potential has been found by Sanyal \cite{sanyal02} under the assumption that the Lagrangian admits strict Noether symmetry. This work was generalized to BI and BIII by Camci \& Kucukakca \cite{camci}. Now we seek the condition for the Lagrangian
(\ref{lag-b13ks}) to admit NGS.

Let us consider a NGS generator
\begin{equation}
{\bf X} = \xi \frac{\partial}{\partial t} + X^i \frac{\partial}{\partial Q^i}, \label{ngs-gen}
\end{equation}
where the configuration space of the Lagrangian is $Q^i = (A, B, \Phi)$  with tangent space  $TQ =(A,B,\Phi,\dot{A},\dot{B},\dot{\Phi})$,  $X^i = (\alpha, \beta, \gamma)$, $i= 1,2,3$; and the components $\xi, \alpha, \beta$ and $\gamma$ are functions of $t, A, B$ and $\Phi$.
The existence of a NGS implies the existence of a vector field ${\bf X}$ as given in (\ref{ngs-gen}), if the Lagrangian $\mathcal{L}(t,Q^i, \dot{Q}^i)$ satisfies
\begin{equation}
{\bf X}^{[1]} \mathcal{L} + \mathcal{L} ( D_t \xi) = D_t f, \label{ngs-eq}
\end{equation}
where ${\bf X}^{[1]}$ is the first prolongation of the NGS generator (\ref{ngs-gen}) in the form
\begin{equation}
{\bf X}^{[1]} = {\bf X}  + \dot{X}^i \frac{\partial}{\partial \dot{Q}^i},
\end{equation}
$f(t,A,B,\Phi)$ is a gauge function, $D_t$ is the total derivative operator with respect to $t$
\begin{equation}
D_t =\frac{\partial}{\partial t} + \dot{Q}^i \frac{\partial}{\partial Q^i},
\end{equation}
and $\dot{X}^i$ is defined as $\dot{X}^i = D_t X^i - \dot{Q}^i D_t \xi$. It is important to give the following first integral to emphasize the significance of NGS: If  ${\bf X}$ is the NGS generator corresponding to the Lagrangian $\mathcal{L}(t,Q^i,\dot{Q}^i)$, then
\begin{equation}
I =-\xi E_{\mathcal{L}} + X^i \frac{\partial \mathcal{L}}{\partial \dot{Q}^i} - f, \label{first-int}
\end{equation}
is the Noether first integral, i.e. the Hamiltonian or a conserved quantity associated with the generator ${\bf X}$. Here, $E_L$ is the energy function defined for any Lagrangian, $E_{\mathcal{L}} = \dot{Q}^i \frac{\partial \mathcal{L}}{\partial \dot{Q}^i} - \mathcal{L}$.

Obviously, the  gauge function $f$ is arbitrary up to an additive constant, and this arbitrariness will be used to simplify expressions in the rest of the paper, whenever possible. Also, the trivial Noether gauge symmetry $\p_t$ is related to the conservation of energy, and gives rise to the Hamiltonian ($E_{\mathcal{L}} = 0$) of the dynamical system.

From the Lagrangian (\ref{lag-b13ks}) the NGS condition (\ref{ngs-eq}) yields the
following set of over-determined system of equations
\begin{equation}
\xi_{A} = 0, \quad \xi_{B} = 0, \quad \xi_{\Phi} = 0,
\end{equation}
\begin{equation}
2 \epsilon B \left( 2 F \beta_t + F' B \gamma_t \right) - G_A = 0,
\end{equation}
\begin{equation}
4 \epsilon \left[ F \left( B \alpha_t + A \beta_t \right) + F' A B \gamma_t \right]- G_B = 0,
\end{equation}
\begin{equation}
2 \epsilon F' \left( B^2 \alpha_t + 2 A B \beta_t \right) + A B^2 \gamma_t - G_{\Phi} = 0,
\end{equation}
\begin{equation}
2 F \beta_ A + B F'\gamma_A =0, \label{neq1}
\end{equation}
\begin{equation}
\alpha + 2B \alpha_B + 2 A \beta_B + A \frac{F'}{F} \left( \gamma + 2 B \gamma_B \right) - A \xi_t = 0, \label{neq2}
\end{equation}
\begin{equation}
\alpha + 2 \frac{A}{B} \beta  +  2 A \gamma_{\Phi} + 4 \epsilon F' \left( \alpha_{\Phi} + 2 \frac{A}{B} \beta_{\Phi} \right)- A \xi_t = 0,\,\, \quad  \label{neq3}
\end{equation}
\begin{equation}
\beta + B \left[ \alpha_A + \beta_B + \frac{F'}{F} \left( \gamma + A \gamma_A + \frac{B}{2} \gamma_B \right) \right]  +  A \beta_A - B \xi_t = 0, \quad \label{neq4}
\end{equation}
\begin{equation}
\frac{F'}{F} \left( 2 \beta + B \alpha_A + B \gamma_{\Phi} + 2 A \beta_A - B \xi_t \right) + \frac{F''}{F} B \gamma  + 2 \beta_{\Phi} + \frac{\epsilon}{2 F} A B \gamma_A = 0, \qquad \label{neq5}
\end{equation}
\begin{equation}
\frac{F'}{F}
\left( \alpha + \frac{A}{B} \beta + \frac{B}{2} \alpha_B + A \beta_B + A
\gamma_{\Phi} - A \xi_t \right)+ \frac{F''}{F} A \gamma + \alpha_{\Phi} +
\frac{A}{B} \beta_{\Phi} + \frac{\epsilon}{4 F} A B \gamma_B = 0, \label{neq6}
\end{equation}
\begin{equation}
 2 \epsilon q F \left( \alpha + \frac{F'}{F} A \gamma + A \xi_t \right) + A B^2 \gamma U'(\Phi)
+  B \left( B \alpha + 2 A \beta  + A B \xi_t \right) U(\Phi) = 0.
\label{neq7}
\end{equation}
Then, selecting the function $F$, the above NGS equations will give the solutions for $\xi, \alpha, \beta, \gamma, f$ and the potential $U(\Phi)$.

For BI, BIII and KS spacetimes the Hessian determinant $W = \Sigma \left| \frac{\partial^2 \mathcal{L}}{\partial \dot{Q}_i \partial \dot{Q}_j} \right|$ is given by
\begin{equation}
W = 16 A B^4 F (3 \epsilon F'^2 - F). \label{hess}
\end{equation}
There exists two cases depending on the Hessian determinant $W$ vanishes or not:
\\ {\bf Case (i)}: If the Lagrangian (\ref{lag-b13ks}) is degenerate, then  the Hessian determinant $W$ vanishes, and therefore the function $F$ is given by
\begin{equation}
F(\Phi) = \frac{\epsilon}{12} \Phi^2.
\label{f-degenerate}
\end{equation}

In this case, the first two of the three main terms of eq.(\ref{cont}) vanish, enabling us to directly determine
\begin{equation}
U(\Phi) = \lambda \Phi^4
\label{u-degenerate}
\end{equation}
whenever (\ref{cont}) applies.

\noindent{\bf Case (ii)}:  If the Lagrangian (\ref{lag-b13ks}) is non-degenerate, then the Hessian determinant $W$ does not vanish. For the form of $F$, we will consider {\bf (ii.a)} $F = F_0 \Phi^2 /12$, where $F_0 \neq \epsilon$, the form motivated by induced gravity; or {\bf (ii.b)} $F(\Phi)=1- \zeta \Phi^2$, which is a more standard form of non-minimally coupled scalar field theory, where $\zeta$ is a constant.

We now present solutions of the NGS equations in the KS, BI and BIII spacetimes for the coupling functions $F(\Phi)$ listed above. The NGS equations also allow determination of $U(\Phi)$ via eq.(\ref{neq7}), but we will classify the solutions according to the $U(\Phi)$ functions from the beginning, in addition to the classification according to the $F(\Phi)$ functions outlined just above. We will show the results in tables, elaborating on only some of them in the text.

\subsection{Bianchi I Spacetime}
\label{b1-spacetime}

We show some of the NGSs in Tables \ref{b1-t1} and \ref{b1-t2}. In order to keep the tables compact, we explicitly present the simplest case {\bf (i.1)} in the text, and then express some results of some other cases in terms of the explicitly presented ones. We also explicitly display some cases, {\bf (i.3), (ii.a.1)} and {\bf (ii.a.4)}, that are too long to fit in the tables.

{\bf Case (i.1)}: $U(\Phi) = 0$. For this case, the \emph{eight} NGSs are given by
\begin{eqnarray}
& & {\bf X}_1 = \p_t, \qquad {\bf X}_2 = -2 A \p_A + B \p_B, \nonumber \\& & {\bf X}_3 = - 2 A \p_A + \Phi \p_\Phi, \qquad {\bf X}_4 = t\p_t + A \p_A, \nonumber \\& & {\bf X}_5 = t^2\p_t + 2 t \left( A \p_A +  B \p_B - \Phi \p_\Phi \right), \nonumber \\& & {\bf X}_6 = 2 A \ln{(B \Phi)} \p_A - B \ln{(A \Phi)} \p_B + \Phi \ln{\left(\frac{A}{B}\right)} \p_\Phi, \qquad \\& & {\bf X}_7 = -\frac{2 A}{B \Phi} \p_A + \frac{1}{\Phi} \p_B + \frac{1}{B} \p_\Phi, \nonumber \\& & {\bf X}_8 = 2 \frac{(A/B)^{\frac{1}{3}}}{\Phi} \p_A - \frac{(B/A)^{\frac{2}{3}}}{\Phi} \p_B + (A^2 B)^{-\frac{1}{3}} \p_{\Phi}, \nonumber
\end{eqnarray}
with the non-vanishing Lie brackets
\begin{eqnarray}
& & [{\bf X}_1,{\bf X}_4]={\bf X}_1 \,\,  \qquad
[{\bf X}_1,{\bf X}_5]=2({\bf X}_2 - {\bf X}_3+{\bf X}_4), \nonumber \\& &
[{\bf X}_2, {\bf X}_6]=2{\bf X}_2-3{\bf X}_3, \quad [{\bf X}_2, {\bf X}_7]=-{\bf X}_7, \nonumber\\& & [{\bf X}_2, {\bf X}_8]={\bf X}_8, \qquad
[{\bf X}_3, {\bf X}_6]={\bf X}_2-2{\bf X}_3, \nonumber \\& &
[{\bf X}_3, {\bf X}_7]= - {\bf X}_7, \quad \,\,
[{\bf X}_3, {\bf X}_8]= \frac{1}{3}{\bf X}_8, \nonumber \\& &
[{\bf X}_4, {\bf X}_5]= {\bf X}_5, \qquad
[{\bf X}_4, {\bf X}_6]= {\bf X}_3-{\bf X}_2, \label{commut-i1} \\& &
[{\bf X}_4, {\bf X}_8] = - \frac{2}{3} {\bf X}_8, \nonumber \\& &
[{\bf X}_6, {\bf X}_7] = \frac{1}{B \Phi}({\bf X}_2 - 3 {\bf X}_3) + \ln{(B \Phi)} {\bf X}_7, \nonumber \\& & [{\bf X}_6, {\bf X}_8] = \frac{3}{(A^2 B)^{1/3} \Phi}({\bf X}_2 - {\bf X}_3)  + \ln{\left( \frac{1}{(A^2 B)^{1/3} \Phi} \right)} {\bf X}_8. \nonumber
\end{eqnarray}
Using (\ref{first-int}), we find the first integrals for those of eight NGSs:
\begin{eqnarray}
& & I_1 = - E_L, \quad
I_2 = \frac{1}{3} A B^2 \Phi^2 \left(\frac{\dot{A}}{A} - \frac{\dot{B}}{B}\right), \label{fint12-b1-i1} \\& & I_3 = \frac{1}{3} A B^2 \Phi^2 \left( \frac{\dot{A}}{A} + \frac{\dot{\Phi}}{\Phi} \right), \label{fint3-b1-i1} \\& &  I_4 = - t E_L + \frac{1}{3} A B^2 \Phi^2 \left( \frac{\dot{B}}{B} + \frac{\dot{\Phi}}{\Phi} \right), \label{fint4-b1-i1}  \\& & I_5 = - t^2 E_L, \label{fint5-b1-i1} \\& & I_6 = \frac{1}{3} A B^2 \Phi^2 \Big{[} -\ln{(B \Phi)} \frac{\dot{A}}{A} +\ln{( A \Phi )}  \frac{\dot{B}}{B} + \ln{(\frac{A }{B})} \frac{\dot{\Phi}}{\Phi} \Big{]}, \label{fint6-b1-i1}  \\& & I_7 = \frac{1}{3} A B \Phi \left( 2 \frac{\dot{A}}{A} + \frac{\dot{B}}{B} + 3 \frac{\dot{\Phi}}{\Phi} \right), \label{fint7-b1-i1} \\& &  I_8 = (A B^5)^{1/3} \Phi \left(  \frac{\dot{B}}{B} + \frac{\dot{\Phi}}{\Phi} \right).  \label{fint8-b1-i1}
\end{eqnarray}

{\bf Case (i.3)}: $U(\Phi) = \lambda \Phi^m$, where $m (\neq 2)$ is a constant. In this case, it follows from the continuity equation (\ref{cont}) that $m = 4$. Then the NGS components are obtained by
\begin{eqnarray}
& & \xi = g(t), \qquad  \alpha = A \dot{g}(t) -\frac{2}{B \Phi^3} \left( c_1 \frac{A}{3 B^2} + \frac{c_2}{A} \right)- 2 c_3 A, \nonumber \\& & \beta = B \dot{g}(t) - \frac{5 c_1}{3 B^2 \Phi^3} -\frac{c_2}{A^2 \Phi^3}  + c_3 B, \, \gamma = - \Phi \dot{g}(t) + \frac{1}{B \Phi^2} \left( \frac{c_1}{B^2}  + \frac{c_2}{A^2} \right), \qquad
\end{eqnarray}
where $g(t)$ is an arbitrary function of $t$. This result is interesting because of that there exists an infinite family of NGSs due to the functional dependence of NGS components.
Thus the \emph{four} NGSs of this case are ${\bf X}^3_1 = {\bf X}_2$ given above and
\begin{eqnarray}
& & {\bf X}^3_2 =- \frac{2}{A B \Phi^3}\p_A - \frac{1}{A^2\Phi^3}\p_B + \frac{1}{A^2 B \Phi^2}\p_\Phi, \nonumber\\ & & {\bf X}^3_3 =   -\frac{2A}{3 B^3 \Phi^3}\p_A -\frac{5}{3B^2\Phi^3}\p_B +\frac{1}{B^3 \Phi^2}\p_\Phi, \\& & {\bf X}^3_4 =  g(t) \p_t + \dot{g} (t) \left(  A \p_A+ B\p_B- \Phi \p_\Phi \right). \nonumber
\end{eqnarray}

\begin{table*}[!ht]
\centering
\caption{ The potential functions and lists of NGSs of cases (i) and (ii) for the BI ($q = 0$) spacetime, where $\lambda$ is a constant. \label{b1-t1}}
\resizebox{17cm}{!} {
\begin{tabular}{|l|c|c|c|}
\hline
 Potential Function & $ Case(i)\, F = \epsilon \Phi^2 / 12 $ & $ Case(ii.a)\, F = F_0 \Phi^2  / 12 $ & $Case(ii.b)\, F = 1 - \zeta \Phi^2 $ \\
\hline
$1.\, U(\Phi) = 0 $ & $ {\bf X}_1, {\bf X}_2, {\bf X}_3, {\bf X}_4, {\bf X}_5, {\bf X}_6, {\bf X}_7, {\bf X}_8 $ & $ {\bf X}_1, {\bf X}_2, {\bf X}_3, {\bf X}_4, {\bf X}_5^{a1}, {\bf X}_6^{a1} $ & $ {\bf X}_1, {\bf X}_2, {\bf X}_4 $ \\

$2.\, U(\Phi) = \lambda $ & $ {\bf X}_1, {\bf X}_2, {\bf X}_3, {\bf X}_6, {\bf X}_5^{2} $ & $ {\bf X}_1, {\bf X}_2, {\bf X}_3^{a2} $ & $ {\bf X}_1, {\bf X}_2 $ \\

$3.\, U(\Phi) = \lambda \Phi^m, m\neq2 $ & $ {\bf X}_2, {\bf X}_2^{3}, {\bf X}_3^{3}, {\bf X}_4^{3} $ & $ {\bf X}_1, {\bf X}_2, {\bf X}_3^{a3} $ & $ {\bf X}_1, {\bf X}_2 $ \\

$4.\, U(\Phi) = \lambda \Phi^2 $ & $ {\bf X}_1, {\bf X}_2, {\bf X}_3, {\bf X}_6, {\bf X}_5^{4} $ & $ {\bf X}_1, {\bf X}_2, {\bf X}_3, {\bf X}_4^{a4}, {\bf X}_5^{a4}, {\bf X}_6^{a4} $ & $ {\bf X}_1, {\bf X}_2 $ \\
\hline
\end{tabular}
}
\end{table*}

\begin{table*}[!ht]
\centering
\caption{ The NGSs and corresponding first integrals for the BI spacetime cases not covered in the text. \label{b1-t2}}
\resizebox{17cm}{!} {
\begin{tabular}{|l|l|l|}
\hline
 Case & NGS & First Integral  \\
\hline
(i.2) & ${\bf X}_1^2 = {\bf X}_1,\quad {\bf X}_2^2 = {\bf X}_2,\quad {\bf X}_3^2 = {\bf X}_3,\quad {\bf X}_4^2 = {\bf X}_6 $& $I_1^2 = I_1, \quad I_2^2 = I_2, \quad I_3^2 = I_3, \quad I_4^2 = I_6$ \\
      & ${\bf X}_5^2 = t \p_t - A \p_A + \Phi \p_\Phi$ & $ I_5^2= \frac{1}{3} A B^2 \Phi^2 \Big{(} \frac{\dot{A}}{A} + \frac{\dot{B}}{B} + 2 \frac{\dot{\Phi}}{\Phi} \Big{)}$ \\
    \hline
(i.4) & ${\bf X}_1^4 = {\bf X}_1,\quad {\bf X}_2^4 = {\bf X}_2,\quad {\bf X}_3^4 = {\bf X}_3,\quad {\bf X}_4^4 = {\bf X}_6$ & $I_1^4 = I_1, \quad I_2^4 = I_2, \quad I_3^4 = I_3, \quad I_4^4 = I_6$ \\
      & ${\bf X}_5^4 = t\p_t - A [4\ln{(B\Phi)}+1]\p_A-B\ln{(B\Phi)}\p_B $ & $I_5^4 = \frac{1}{3} A (B \Phi)^2 \left[  \ln{(B \Phi)} \left( 2 \frac{\dot{A}}{A} +  \frac{\dot{B}}{B} + 3 \frac{\dot{\Phi}}{\Phi} \right) - \left( \frac{\dot{B}}{B} + \frac{\dot{\Phi}}{\Phi} \right)  \right]  $  \\
      & $\qquad \,\,\, + 3\Phi\ln{(B\Phi)}\p_\Phi$ &  \\
    \hline
(ii.a.2) & $ {\bf X}_1^{a2} = {\bf X}_1,\quad {\bf X}_2^{a2} = {\bf X}_2,\quad {\bf X}_3^{a2} = {\bf X}_5^2 $ & $ I_1^{a2} = I_1, \quad I_2^{a2} = I_2, \quad I_3^{a2} = I_5^2$  \\
    \hline
(ii.a.3) & $ {\bf X}_1^{a3}={\bf X}_1,\quad {\bf X}_2^{a3} = {\bf X}_2 $ & $ I_1^{a3} = I_1, \quad I_2^{a3} = I_2 $ \\
         & $ {\bf X}_3^{a3} = t\p_t + \frac{(m+2)}{(m-2)} A \p_A -\frac{2}{(m-2)} \Phi \p_\Phi $ & $ I_3^{a3} = \frac{k_3}{3 (m - 2)} A (B \Phi)^2 \left[ m \left( \frac{\dot{B}}{B} + \frac{\dot{\Phi}}{\Phi} \right) - 2 \left( \frac{\dot{A}}{A} +  \frac{\dot{B}}{B} - \frac{\dot{\Phi}}{\Phi} \right) - \frac{6}{k_3} \frac{\dot{\Phi}}{\Phi} \right]   $  \\
    \hline
(ii.b.1) & $ {\bf X}_1^{b1} = {\bf X}_1,\quad {\bf X}_2^{b1} = {\bf X}_2,\quad {\bf X}_3^{b1} = {\bf X}_4 $ & $ I_1^{b1} = I_1, \quad I_2^{b1} = I_2, \quad I_3^{b1} = I_4 $ \\
    \hline
(ii.b.2) & $ {\bf X}_1^{b2} = {\bf X}_1,\quad {\bf X}_2^{b2} = {\bf X}_2 $ & $ I_1^{b2} = I_1, \quad I_2^{b2} = I_2 $ \\
	\hline
(ii.b.3) & $ {\bf X}_1^{b3} = {\bf X}_1,\quad {\bf X}_2^{b3} = {\bf X}_2 $ & $ I_1^{b3} = I_1, \quad I_2^{b3} = I_2 $ \\
	\hline
(ii.b.4) & $ {\bf X}_1^{b4} = {\bf X}_1,\quad {\bf X}_2^{b4} = {\bf X}_2 $ & $ I_1^{b4} = I_1, \quad I_2^{b4} = I_2 $ \\		
\hline
\end{tabular}
}
\end{table*}

The non-zero Lie brackets due to the NGS generators ${\bf X}_3^3$ and ${\bf X}_4^3$ are
\begin{eqnarray}
& & [{\bf X}_2, {\bf X}_3^3] = -3{\bf X}_3^3, \quad
[{\bf X}_2, {\bf X}_2^3] = 3 {\bf X}_2^3.
\end{eqnarray}
The first integrals for the NGSs of this case are given by
\begin{eqnarray}
& & I^3_1 = \frac{1}{3} A B^2 \Phi^2 \left(\frac{\dot{A}}{A} - \frac{\dot{B}}{B}\right), \quad I^3_2 = -\frac{B}{3 A \Phi} \left( \frac{\dot{B}}{B} + \frac{\dot{\Phi}}{\Phi} \right), \nonumber  \\& & I^3_3 = -\frac{A}{9 B \Phi } \left( 2 \frac{\dot{A}}{A} + \frac{\dot{B}}{B} + 3 \frac{\dot{\Phi}}{\Phi} \right), \qquad  I^3_4 = -g(t) E_L. \label{fint-b1-i-3}
\end{eqnarray}

{\bf Case (ii.a.1)}: $U(\Phi) = 0$. For this case, the components of NGS generators, and the gauge function are found as
\begin{eqnarray}
& & \xi =  c_1 \frac{t^2}{2}+ c_2 t + c_3, \quad
\alpha = A \left( c_1 \frac{ k_4}{k_1} t  + c_2 - 2 c_4 \right)  + 2 c_5 A \ln{(B \Phi^{\frac{k_4}{k_3}}}) - 2 c_6, \,\, \nonumber \\& & \beta = c_1 \frac{_5}{k_1} B t + c_5 B \ln{(\frac{\Phi^{\frac{k_4}{k_3}}}{A}}) + c_6 B , \, \gamma = c_1 \frac{k_3}{k_1} t \Phi + c_4 \Phi + c_5 \Phi \ln{(A/B)}, \qquad \\& & f = c_1 \frac{k_2 k_3}{k_1} A B^2 \Phi^2, \nonumber
\end{eqnarray}
where $k_1 = 8 \epsilon F_0 - 9, k_2 = \epsilon F_0 - 1, k_3 = \epsilon F_0$ and $k_4 = 2 \epsilon F_0 - 3$. It is easily seen that $k_2$ and $k_3$ are non-zero in this case,
but $k_1$ or $k_4$ could be. Hence, we need to consider subcases where they are both nonzero, vs. where one or the other vanishes. The latter singular ones cannot be obtained as special cases of the above solutions, they have to be trated from scratch.

\noindent \underline{Subcase ii.a.1.1.} $k_1 \neq 0$, $k_4 \neq 0$. In this subcase, the number of NGSs is \emph{six} which are ${\bf X}_1^{a1} = {\bf X}_1, {\bf X}_2^{a1} = {\bf X}_2, {\bf X}_3^{a1} = {\bf X}_3, {\bf X}_4^{a1} = {\bf X}_4$ and
\begin{eqnarray}
& & {\bf X}_5^{a1} = t^2 \p_t + \frac{2t}{k_1} \left[ k_4 ( A \p_A + B \p_B ) + k_3 \Phi  \p_\Phi \right], \quad \\ & & {\bf X}_6^{a1} = 2 A \ln{ \left( B \Phi^{\frac{k_4}{k_3}} \right) } \p_A + B \ln{ \left( \frac{\Phi^{\frac{k_4}{k_3}}}{A} \right) } \p_B  + \Phi \ln{(A/B)} \p_\Phi.
\end{eqnarray}
The non-vanishing Lie brackets due to ${\bf X}_5^{a1} $ and ${\bf X}_6^{a1}$ are
\begin{eqnarray}
& & [{\bf X}_1,{\bf X}_5^{a1}] = \frac{k_4}{k_1}{\bf X}_2 + \frac{k_3}{k_1}{\bf X}_3 + {\bf X}_4, \,\,  [{\bf X}_2, {\bf X}_6^{a1}] = 2 {\bf X}_2 - 3 {\bf X}_3,  \nonumber \\& & [{\bf X}_3, {\bf X}_6^{a1}] = - 2 {\bf X}_3 - c {\bf X}_2, \,\,  [{\bf X}_4 , {\bf X}_5^{a1}] = {\bf X}_5^{a1}, \,\,  [{\bf X}_4, {\bf X}_6^{a1}] = {\bf X}_4 - {\bf X}_6^{a1}. \qquad
\end{eqnarray}
where $c  = (3 \epsilon - 4 F_0)/ F_0 $. The first integrals of the generators ${\bf X}_5^{a1}$ and ${\bf X}_6^{a1}$ are
\begin{eqnarray}
& & I_5 = \frac{2 k_2 k_3}{k_1} A B^2 \Phi^2 \Big{[} t \Big{(}\frac{\dot{A}}{A} + 2 \frac{\dot{B}}{B} + 2 \frac{\dot{\Phi}}{\Phi} \Big{)} - 1 \Big{]}, \\& & I_6 = \frac{2 k_3 }{3} t A B^2 \Phi^2 \Big{[} \frac{1}{2} \ln{(B^{-1} \Phi^{\ell})} \frac{\dot{A}}{A} + \ln{( \Phi^{\frac{3 \ell}{2}})} \frac{\dot{B}}{B} + \ln{(B \Phi^4 / A)^{\ell/2}} \frac{\dot{\Phi}}{\Phi} \Big{]}, \quad
\end{eqnarray}
where $ \ell = k_4 / k_3 $.

\noindent \underline{Subcase ii.a.1.2.}  $ F_0= 9 \epsilon /8$, i.e. $k_1 =0$. There are {\emph{seven} NGSs for this subcase, which are ${\bf Y}_1 = {\bf X}_1, {\bf Y}_2 = {\bf X}_2, {\bf Y}_3 = {\bf X}_3, {\bf Y}_4 = {\bf X}_4$ and
\begin{eqnarray}
& & {\bf Y}_5 =- 8 t \left[ \frac{2}{3} \left( A \p_A + B \p_B \right) - \Phi \p_\Phi \right], \\ & & {\bf Y}_6 = -\frac{2A}{3} [\ln{( B \Phi ) }^{2} +1]\p_A -\frac{B}{3} [\ln{ (B\Phi^{2} ) } ]\p_B +\Phi \ln{( B \Phi^{\frac{4}{3}} ) } \p_\Phi ,  \\ & &
{\bf Y}_7 = \frac{2A}{3} [\ln{ (B)} - 1]\p_A -\frac{B}{3} [\ln{ (B A^{3}\Phi^{4} ) } ]\p_B  + \Phi \ln{ ( A\Phi^{\frac{4}{3}} )  } \p_\Phi,
\end{eqnarray}
where we have a nonzero gauge function for ${\bf Y}_5$ as $f= A B^2 \Phi^2$. The non-zero Lie brackets for this subcase are
\begin{eqnarray}
& & [{\bf Y}_1, {\bf Y}_5] = -\frac{16}{3} {\bf Y}_2 + 8 {\bf Y}_3, \,\, [{\bf Y}_2, {\bf Y}_6] = -\frac{1}{3} {\bf Y}_2 + {\bf Y}_3, \nonumber\\& & [{\bf Y}_2, {\bf Y}_7] = \frac{5}{3} {\bf Y}_2 - 2 {\bf Y}_3,  \quad [{\bf Y}_3, {\bf Y}_6] = -\frac{2}{3}( {\bf Y}_2 - 2 {\bf Y}_3 ), \nonumber \\& & [{\bf Y}_3, {\bf Y}_7] = \frac{2}{3}( {\bf Y}_2 - {\bf Y}_3 ), \quad [{\bf Y}_4, {\bf Y}_5] = {\bf Y}_5, \\& & [{\bf Y}_4, {\bf Y}_7] = - {\bf Y}_2 + {\bf Y}_3,  \quad [{\bf Y}_5, {\bf Y}_6] = \frac{2}{3} {\bf Y}_5, \nonumber\\& & [{\bf Y}_5, {\bf Y}_7] = \frac{2}{3} {\bf Y}_5, \quad [{\bf Y}_6, {\bf Y}_7] = \frac{2}{3} ( {\bf Y}_2 - {\bf Y}_3 ) + \frac{1}{3} ( {\bf Y}_6 - {\bf Y}_7 ). \nonumber
\end{eqnarray}
The first integrals related to ${\bf Y}_5, {\bf Y}_5$ and ${\bf Y}_7$ are
\begin{eqnarray}
& & I_5 =  A B^2 \Phi^2 \left[ t \Big{(}\frac{\dot{A}}{A}+2\frac{\dot{B}}{B}+2\frac{\dot{\Phi}}{\Phi}\Big{)} - 1 \right], \\& & I_6 = \frac{ 1}{4} A B^2 \Phi^2 \Big{[} \ln{(B \Phi)} \frac{\dot{A}}{A} + \Big{(}\ln{(B^{\frac{1}{2}} \Phi)} -1\Big{)}\frac{\dot{B}}{B}  + \Big{(}\ln{(B\Phi^{\frac{4}{3}})}-1\Big{)}\frac{\dot{\Phi}}{\Phi} \Big{]}, \qquad  \\& &
I_7 =\frac{1}{8}  \Big{[} - \ln{(B)} \frac{\dot{A}}{A} + (\ln{(A^{3}B\Phi^{4})} -2)\frac{\dot{B}}{B} + (\ln{(A^{3}\Phi^{-4})}-2)\frac{\dot{\Phi}}{\Phi}\Big{]}.
\end{eqnarray}

\noindent \underline{Subcase ii.a.1.3.}  $F_0= 3 \epsilon /2$, i.e. $k_4 =0$. There are \emph{six} NGSs for this subcase:  ${\bf Y}_1 = {\bf X}_1, {\bf Y}_2 = {\bf X}_2, {\bf Y}_3 = {\bf X}_3, {\bf Y}_4 = {\bf X}_4$ and
\begin{eqnarray}
 & &  {\bf Y}_5 = \frac{t^2}{2} \p_t + \frac{t}{2} \Phi \p_\Phi,  \\ & &
 {\bf Y}_6 = 2A\ln{(B)}\p_A-B\ln{(A)}\p_B-\Phi\ln{(\frac{A}{B})}\p_\Phi, \qquad
\end{eqnarray}
with a nonzero gauge function $f =\frac{1}{4} A B^2 \Phi^2$ for ${\bf Y}_5$. The non-zero Lie brackets of the above vector fields yield
\begin{eqnarray}
& & [{\bf Y}_1, {\bf Y}_5] = \frac{1}{2} {\bf Y}_3 + {\bf Y}_4, \, [{\bf Y}_2, {\bf Y}_6] = 2 {\bf Y}_2 - 3 {\bf Y}_3, \qquad \quad \nonumber \\& & [{\bf Y}_3, {\bf Y}_6] = 2 ( {\bf Y}_2 - {\bf Y}_3 ), \, [{\bf Y}_4, {\bf Y}_5] = {\bf Y}_5, \\& & [{\bf Y}_4, {\bf Y}_6] = - {\bf Y}_2 + {\bf Y}_3. \nonumber
\end{eqnarray}
The first integrals of ${\bf Y}_5$ and ${\bf Y}_6$ are
\begin{eqnarray}
& & I_5 =  \frac{1}{4} A B^2 \Phi^2 \left[ t \Big{(}\frac{\dot{A}}{A}+2\frac{\dot{B}}{B}+ 2 \frac{\dot{\Phi}}{\Phi}\Big{)} -1 \right], \\& &
I_6= A B^2 \Phi^2 \Big{[} \ln{(\frac{B^{1/2}}{A})}\frac{\dot{A}}{A}+\ln{(\frac{B^{2}}{A^{3/2}})}\frac{\dot{B}}{B} + 2 \ln{( B / A) } \frac{\dot{\Phi}}{\Phi} \Big{]}.
\end{eqnarray}

{\bf Case (ii.a.4)}: $U(\Phi) = \lambda \Phi^2$. For this case, we find the following NGS components and gauge function:
\begin{eqnarray}
& & \xi = c_1 +  c_2 \sin{ \left( a t \right) } + c_3 \cos{ \left( a t \right) }, \nonumber \\& & \alpha = \frac{a k_4}{k_1} A \left[ c_2 \cos{ \left( a t \right) } - c_3 \sin{ \left( a t \right) } \right] - c_5 \frac{2 A }{k_2 k_3}\ln{(B\phi)}- 2 (c_4 + c_6) A, \nonumber \\& & \beta = \frac{a k_4}{k_1} B \left[ c_2 \cos{ \left( a t \right) } - c_3 \sin{ \left( a t \right) } \right]  - c_5 B \ln{(A\phi^{-k_4 / F_0})} + c_6 B, \\& & \gamma = \frac{a k_3}{k_1} \Phi \left[ c_2 \cos{ \left( a t \right) } - c_3 \sin{ \left( a t \right) } \right]   + c_4 \Phi + c_5 \Phi \ln{( A / B )}, \nonumber \\& &  f = - 2 \lambda A B \Phi^2 \left[ c_2 \sin{ \left( a t \right) }  + c_3 \cos{ \left( a t \right) } \right]. \nonumber
\end{eqnarray}
where $a = \sqrt{2 \lambda k_1 / k_2 k_3}, k_1 = 8 \epsilon F_0 - 9, k_2 = \epsilon F_0 - 1, k_3 = \epsilon F_0$ and $k_4 = 2 \epsilon F_0 - 3$. Here we observe that $k_2$ is different from zero because $F_0 \neq \epsilon$, and it is clear that $k_3 \neq 0$. Therefore, again we need to consider subcases analogous to the Case (ii.a.1).

\noindent \underline{Subcase ii.a.4.1.} $k_1 \neq 0$,  $k_4 \neq 0$. The NGSs are obtained as ${\bf X}_1^{a4} = {\bf X}_1, {\bf X}_2^{a4} = {\bf X}_2, {\bf X}_3^{a4} = {\bf X}_3$ and
\begin{eqnarray}
& & {\bf X}_4^{a4} = 2 A \ln{(B \phi^{-k_4/F_0})} \p_A - B \ln{(A \phi^{-k_4/F_0})} \p_B + \Phi \ln{( A / B )} \p_\Phi, \nonumber \\& & {\bf X}_5^{a4} = \frac{a}{k_1} \cos{\left( a t \right)} \Big{[} k_4 ( A \p_A + B \p_B ) + k_3 \Phi \p_\Phi \Big{]}  + \sin{ \left( a t \right) } \p_t , \\& & {\bf X}_6^{a4} =  - \frac{a}{k_1 } \sin{\left( a t \right)}  \Big{[} k_4 ( A \p_A + B \p_B ) + k_3 \Phi \p_\Phi \Big{]} + \cos{ \left( a t \right) } \p_t, \nonumber
\end{eqnarray}
where $f = - 2 \lambda A B \Phi^2 \sin{ \left( a t \right) }$ and $f = - 2 \lambda A B \Phi^2 \cos{ \left( a t \right) }$ for ${\bf X}_5^{a4}$ and ${\bf X}_6^{a4}$, respectively.
The non-zero Lie brackets for this case are
\begin{eqnarray}
& & [{\bf X}_1, {\bf X}_5^{a4}] = a {\bf X}_6^{a4}, \qquad  \,\, [{\bf X}_1, {\bf X}_6^{a4}] = - a {\bf X}_5^{a4}, \\& & [{\bf X}_2, {\bf X}_4^{a4}] = - 3 {\bf X}_4 + 2 {\bf X}_2, \\& & [{\bf X}_3, {\bf X}_4^{a4}] = 2 {\bf X}_3 + b {\bf X}_2 , \,\, [{\bf X}_5^{a4}, {\bf X}_6^{a4}] = - a {\bf X}_1.
\end{eqnarray}
where $b = (3 \epsilon - 4 F_0) / F_0$.
Then, the first integrals read
\begin{eqnarray}
& & I_4 = \frac{k_3 }{3} A B \Phi^2 \Big{[} \ln{( \Phi^{\frac{k_4}{F_0}} /B )} \frac{\dot{A}}{A} + \ln{( B \Phi^{-\frac{k_4}{F_0}}) } \frac{\dot{B}}{B} + \frac{k_4}{k_3} \ln{( B / A )} \frac{\dot{\Phi}}{\Phi} \Big{]}, \\& & I_5 = \frac{3 a k_2 k_3 }{k_1} \cos{(a t)} A B^2 \Phi^2 \Big{[}   \frac{\dot{A}}{A} + 2 \frac{\dot{B}}{B} + 2 \frac{\dot{\Phi}}{\Phi} + \frac{a \tan{(a t)}}{B}  \Big{]}, \\& & I_6 = -\frac{3 a k_2 k_3}{k_1} \sin{(a t)} A B^2 \Phi^2 \Big{[} \frac{\dot{A}}{A} + 2 \frac{\dot{B}}{B} + 2 \frac{\dot{\Phi}}{\Phi} - \frac{ a \cot{(a t)}}{B}  \Big{]}.
\end{eqnarray}

\noindent \underline{Subcase ii.a.4.2.} $F_0= 9 \epsilon / 8$, i.e. $k_1= 0$. The \emph{six} NGSs are found as ${\bf Y}_1 = {\bf X}_1, {\bf Y}_2 = {\bf X}_2,{\bf Y}_3 = {\bf X}_3 $ and
\begin{equation}
{\bf Y}_4 = t \p_t + \frac{1}{3} (16 \lambda t^2 + 3) A \p_A + 8 \lambda t^2 (\frac{2 }{3} B \p_B - \Phi \p_\Phi),
\end{equation}
\begin{equation}
{\bf Y}_5 = -\frac{2}{3}t ( A \p_A + B \p_B )+ \phi \p_\phi
\end{equation}
\begin{equation}
{\bf Y}_6 = - 2A \ln{(B \phi^{2/3})} \p_A - B \ln{(A \phi^{2/3})} \p_B + \phi\ln{(\frac{A}{B})} \p_\phi,
\end{equation}
\begin{eqnarray}
& & {\bf Y}_7 = \frac{1}{6} \Big{[}-8 \lambda t^2 + 3 \ln{(B)} - 3 \Big{]} A \p_A  - \frac{1}{12} \Big{[} 16 \lambda t^2 + 3 \ln{(A^3 B \Phi^4)} \Big{]} B \p_B  \nonumber \\& & \qquad \quad  + \frac{1}{4} \Big{[} 8 \lambda t^2 + \ln{(A^3 \Phi^4)} \Big{]} \Phi \p_\Phi,
\end{eqnarray}
where the corresponding non-zero Lie brackets for this subcase are
\begin{eqnarray}
& & [{\bf Y}_1, {\bf Y}_4] = {\bf Y}_1 - 16 \lambda {\bf Y}_5, \,\,\,\, [{\bf Y}_1, {\bf Y}_5] = -\frac{2}{3} {\bf Y}_2 + {\bf Y}_3, \nonumber \\& & [{\bf Y}_1, {\bf Y}_7] = 4 \lambda {\bf Y}_5, \,\,\,\,\quad \qquad  [{\bf Y}_2, {\bf Y}_6] = -2 {\bf Y}_2 + 3 {\bf Y}_3, \nonumber \\& & [{\bf Y}_2, {\bf Y}_7] = \frac{5}{4} {\bf Y}_2 - \frac{3}{2} {\bf Y}_3, \,\,\,\,  [{\bf Y}_3, {\bf Y}_6] = -\frac{4}{3} {\bf Y}_2 + 2 {\bf Y}_2, \nonumber \\& & [{\bf Y}_3, {\bf Y}_7] = -\frac{1}{2} ( {\bf Y}_2 - {\bf Y}_3 ), \,\,\,\, [{\bf Y}_4, {\bf Y}_5] = {\bf Y}_5,  \\& & [{\bf Y}_4, {\bf Y}_6] = {\bf Y}_2 - {\bf Y}_3,  \,\quad \qquad [{\bf Y}_4, {\bf Y}_7] = -\frac{3}{4} ( {\bf Y}_2 - {\bf Y}_3 ), \nonumber  \\& & [{\bf Y}_5, {\bf Y}_7] = \frac{1}{2} {\bf Y}_5, \,\,\,\,\,\qquad \qquad [{\bf Y}_6, {\bf Y}_7] = \frac{1}{2} ( {\bf Y}_2 - {\bf Y}_3 ) + \frac{1}{4} {\bf Y}_6. \nonumber
\end{eqnarray}
The first integrals of the six NGSs are given by
\begin{eqnarray}
& & I_4 = \frac{1}{8} A B \Phi^2 \Big{[} - 8 \lambda t^2 \frac{\dot{A}}{A} + (3 -16 \lambda t^2) \frac{\dot{B}}{B}  + (40 \lambda t^2 + 3) \frac{\dot{\Phi}}{\Phi} + \frac{4}{3} \lambda t \Big{]}, \\& & I_5 = \frac{1}{8} A B \Phi^2 \Big{[}  (3 - 2t) \frac{\dot{A}}{A} + (6 - 4t) \frac{\dot{B}}{B}   + (8 - 6t) \frac{\dot{\Phi}}{\Phi} + 1 \Big{]}, \\& & I_6 = \frac{3}{8} A B \Phi^2 \Big{[} -\ln{\Big{(} B \Phi^{2/3} \Big{)}}\frac{\dot{A}}{A} + \ln{\Big{(}\frac{A}{(B \Phi)^2}\Big{)}}\frac{\dot{B}}{B} + \ln{\Big{(}\frac{A^{2/3}}{B^{14/3} \Phi^{8/3}}\Big{)}}\frac{\dot{\Phi}}{\Phi} \Big{]}, \qquad \\& & I_7 = \frac{1}{32} A B \Phi^2 \Big{[} (8 \lambda t^2 - 3 \ln{B}) \frac{\dot{A}}{A}  + (16 \lambda t^2 + 3 \ln{(A^3 B \Phi^4 - 6)}) \frac{\dot{B}}{B}  \nonumber \\& & \qquad \quad  + 2(8 \lambda t^2 + \ln{(A^3 \Phi^4 - 3)}) \frac{\dot{\Phi}}{\Phi} + 16 \lambda t  \Big{]}.
\end{eqnarray}

\noindent \underline{Subcase ii.a.4.3.} $k_4= 0$, i.e. $F_0= 3\epsilon / 2$. There are \emph{six} NGSs which are ${\bf Y}_1 = {\bf X}_1, {\bf Y}_2 = {\bf X}_2 , {\bf Y}_3 = {\bf X}_3$ and
\begin{eqnarray}
& & {\bf Y}_4 = 2A\ln{(B)}\p_A-B\ln{(A)}\p_B-\Phi\ln{(\frac{A}{B})}\p_\Phi, \nonumber \\& &
{\bf Y}_5 = sin{2\sqrt{2\lambda}t}\p_t+\Phi \sqrt{2\lambda} \cos{2\sqrt{2\lambda}t}\p_\Phi, \\&&
{\bf Y}_6 = \cos{2\sqrt{2\lambda}t}\p_t-\Phi \sqrt{2\lambda} \sin{2\sqrt{2\lambda}t}\p_\Phi, \nonumber
\end{eqnarray}
with the non-zero Lie brackets
\begin{eqnarray}
& & [{\bf Y}_1, {\bf Y}_5] = 2 \sqrt{2 \lambda}{\bf Y}_6, \quad  [{\bf Y}_1, {\bf Y}_6] = -2 \sqrt{2 \lambda}{\bf Y}_5, \nonumber \\& & [{\bf Y}_2, {\bf Y}_4] = 2 {\bf Y}_2 - 3 {\bf Y}_3,  [{\bf Y}_3, {\bf Y}_4] = 2 {\bf Y}_2 - 2 {\bf Y}_3, \qquad \\& & [{\bf Y}_5, {\bf Y}_6] = -2 \sqrt{2 \lambda}{\bf Y}_1. \nonumber
\end{eqnarray}
The first integrals are
\begin{eqnarray}
& & I_4 = \frac{1}{2} A B \Phi^2 \Big{[} \ln{(\frac{1}{B})} \frac{\dot{A}}{A} + \ln{(A)} \frac{\dot{B}}{B} \Big{]}, \\& & I_5 = \frac{1}{2} A B \Phi^2 \Big{[} \sqrt{2 \lambda} \cos{(2 \sqrt{2 \lambda} t)} \Big{(} \frac{\dot{A}}{A} + 2 \frac{\dot{B}}{B} + 2 \frac{\dot{\Phi}}{\Phi} \Big{)}  + 4 \sin{(2 \sqrt{2 \lambda} t)} \Big{]}, \\& & I_6 = - \frac{1}{2} A B \Phi^2 \Big{[} \sqrt{2 \lambda} \sin{(2 \sqrt{2 \lambda} t)} \Big{(} \frac{\dot{A}}{A} + 2 \frac{\dot{B}}{B} + 2 \frac{\dot{\Phi}}{\Phi} \Big{)} - 4 \cos{(2 \sqrt{2 \lambda} t)} \Big{]}. \quad
\end{eqnarray}

\subsection{Bianchi III and Kantowski-Sachs Spacetimes}
\label{b3ks-spacetime}

The BIII ($q=-1$) and KS ($q=1$) spacetimes, not being as simple as the BI spacetime, allows fewer symmetries. We obtain their NGSs for the same cases, except case (i.3), as for the BI spacetime, showing the results explicitly in Tables \ref{b3ks-t1} and \ref{b3ks-t2}.

{\bf Case (i.3)}: $U(\Phi) = \lambda \Phi^m$, where $m (\neq 2)$ is a constant. We can find solutions only for $m=4$, which gives the following NGS components:
\begin{eqnarray}
& & \xi = g(t), \quad \alpha = A \dot{g}(t) - c_2 \mu_1 + c_3 \mu_2 - 2 c_1 \mu_3, \nonumber \\& & \beta = B \dot{g}(t) - c_2 \mu_4 - c_3 \mu_5 - c_1 \mu_6, \quad \gamma = - \Phi \dot{g}(t) + c_2 \mu_7 + c_3 \mu_8 + c_1 \mu_9, \qquad
\end{eqnarray}
where we have defined $\mu_1, \ldots, \mu_9$ as
\begin{equation}
\mu_1=\frac{2 A (6 \lambda B^2 \Phi^2 - q)}{B \Phi (6 \lambda B^2 \Phi^2 + q)^2}, \,\,  \mu_2=\frac{A [8 \lambda^2 B^4 \Phi^4 + (2 \lambda B^2 \Phi^2 + q)^2]}{q (6 \lambda B^2 \Phi^2 + q)^2}, \,\, \mu_3=(A B \Phi^3)^{-1}
\end{equation}
for $\alpha$,
\begin{equation}
\mu_4=\frac{(30 \lambda B^2 \Phi^2 + q)}{\Phi (6 \lambda B^2 \Phi^2 + q)^2}, \quad \mu_5=\frac{\lambda B^3 \Phi^2 (6 \lambda B^2 \Phi^2 + 5 q )}{q (6 \lambda B^2 \Phi^2 + q)^2},\quad \mu_6=(A^2 \Phi^3)^{-1}
\end{equation}
for $\beta$, and
\begin{equation}
\mu_7=\frac{(18 \lambda B^2 \Phi^2 - q)}{B (6 \lambda B^2 \Phi^2 + q)^2}, \quad  \mu_8=\frac{\Phi (2 \lambda B^2 \Phi^2 - q )}{2 (6 \lambda B^2 \Phi^2 + q)^2}, \quad \mu_9=(A^2 B \Phi^2)^{-1}
\end{equation}
for $\gamma$.

\begin{table*}[!ht]
\centering
\caption{ The potential functions and list of NGSs of cases (i) and (ii) for the BIII and KS spacetimes, where $F_0 \neq \epsilon$. \label{b3ks-t1}}
\resizebox{17cm}{!} {
\begin{tabular}{|l|c|c|c|}
\hline
 Potential Function & Case(i) $F = \epsilon \Phi^2 / 12 $ &  Case(ii.a) $F =  F_0 \Phi^2  / 12 $ & Case(ii.b)$\, F = 1 - \zeta \Phi^2 $ \\[1mm]
\hline
$1.\, U(\Phi) = 0 $ & $ {\bf X}_1, {\bf X}_3, {\bf X}_7, {\bf X}_4^{1} $ & $ {\bf X}_1, {\bf X}_3, {\bf X}_3^{a1} $ & $ {\bf X}_1, {\bf X}_4^1 $ \\

$2.\, U(\Phi) = \lambda $ & $ {\bf X}_1, {\bf X}_7, {\bf X}_3^{2} $ & $ {\bf X}_1, {\bf X}_2^{a2} $ & $ {\bf X}_1 $ \\

$3.\, U(\Phi) = \lambda \Phi^m, m\neq2 $ & $ {\bf X}_1^{3}, {\bf X}_2^{3}, {\bf X}_3^{3}, {\bf X}_4^{3} $ & $ {\bf X}_1, {\bf X}_2^{a3} $ & $ {\bf X}_1 $ \\

$4.\, U(\Phi) = \lambda \Phi^2 $ & $ {\bf X}_1, {\bf X}_3 $ & $ {\bf X}_1, {\bf X}_3$  & $ {\bf X}_1 $ \\
\hline
\end{tabular}
}
\end{table*}

\begin{table*}[!ht]
\centering
\caption{ The NGSs and corresponding first integrals for the BIII and KS spacetimes. The case (1.3) is treated in the text. \label{b3ks-t2}}
\resizebox{17cm}{!} {
\begin{tabular}{|l|l|l|}
\hline
 Case & NGS & First Integral  \\
\hline
(i.1) & ${\bf X}_1^1 = {\bf X}_1, \quad {\bf X}_2^1 = {\bf X}_3, \quad {\bf X}_3^1 = {\bf X}_7,$ & $I_1^1 = I_1, \quad I_2^1 = I_3, \quad I_3^1 = I_7, $ \\
      & ${\bf X}_4^1 = t \p_t - A \p_A + B \p_B$ & $I_4^1 = I_3 $  \\
      \hline
(i.2) & ${\bf X}_1^2 = {\bf X}_1, \quad {\bf X}_2^2 = {\bf X}_7,$ & $I_1^2 = I_1, \quad I_2^2 = I_7,$ \\
      & ${\bf X}_3^2 = t \p_t - 3 A \p_A + B \p_B + \Phi \p_\Phi$ & $I_3^2 = 2 I_3$  \\
      \hline
(i.4) & ${\bf X}_1^4 = {\bf X}_1, \quad {\bf X}_2^4 = {\bf X}_3$ & $I_1^4 = I_1, \quad I_2^4 = I_3$ \\
      \hline
(ii.a.1) & ${\bf X}_1^{a1} = {\bf X}_1, \quad {\bf X}_2^{a1} = {\bf X}_3, \quad {\bf X}_3^{a1} = {\bf X}_4^1$ & $I_1^{a1} = I_1, \quad I_2^{a1} = I_3, \quad I_3^{a1} = I_3$ \\
	\hline
(ii.a.2) & ${\bf X}_1^{a2} = {\bf X}_1, \quad {\bf X}_2^{a2} = {\bf X}_3^2$ & $I_1^{a2} = I_1, \quad I_2^{a2} = 2 I_3$ \\
		\hline
(ii.a.3) & ${\bf X}_1^{a3} = {\bf X}_1,$ & $I_1^{a3} = I_1,$ \\
       &  ${\bf X}_2^{a3} =  t \p_t - \frac{(m - 6)}{(m - 2)} A \p_A + B \p_B - \frac{2}{(m-2)} \Phi \p_\Phi$  & $I_2^{a3} = \frac{F_0}{3 \epsilon (m - 2)}  A B^2 \Phi^2 \left[ (m - 4) \frac{\dot{A}}{A} + \left( m + 2 - \frac{6 \epsilon}{F_0} \right) \frac{\dot{\Phi}}{\Phi} \right]$ \\
	\hline	
(ii.a.4) & ${\bf X}_1^{a4} = {\bf X}_1, \quad {\bf X}_2^{a4} = {\bf X}_3$ & $I_1^{a4} = I_1, \quad I_2^{a4} = I_3$ \\
	\hline
(ii.b.1) & ${\bf X}_1^{b1} = {\bf X}_1, \quad {\bf X}_2^{b1} = {\bf X}_4^1$ & $I_1^{b1} = I_1, \quad I_2^{b1} = I_3$ \\
	\hline
(ii.b.2) & ${\bf X}_1^{b2} = {\bf X}_1$ & $I_1^{b2} = I_1$ \\
	\hline
(ii.b.3) & ${\bf X}_1^{b3} = {\bf X}_1$ & $I_1^{b3} = I_1$ \\
	\hline
(ii.b.4) & ${\bf X}_1^{b4} = {\bf X}_1$ & $I_1^{b4} = I_1$ \\

\hline
\end{tabular}
}
\end{table*}

Then, the \emph{four} NGSs have the form
\begin{eqnarray}
& & {\bf X}^3_1 = - 2 \mu_3 \p_A - \mu_6 \p_B + \mu_9 \p_\Phi, \, \, {\bf X}^3_2 = - \mu_1 \p_A - \mu_4 \p_B + \mu_7 \p_\Phi, \nonumber \\& & {\bf X}^3_3 = \mu_2 \p_A - \mu_5 \p_B + \mu_8 \p_\Phi, \,\, {\bf X}^3_4 = g(t) \p_t + \dot{g}(t) \left[ A \p_A + B \p_B - \Phi \p_\Phi \right], \qquad
\end{eqnarray}
which give rise to the non-zero Lie brackets
\begin{eqnarray}
& & [{\bf X}^3_1,{\bf X}^3_3] = \frac{1}{2 q} {\bf X}^3_1, \qquad [{\bf X}^3_2,{\bf X}^3_3] = - \frac{1}{2 q} {\bf X}^3_2.
\end{eqnarray}
The first integrals of the NGS generators ${\bf X}^3_1, {\bf X}^3_2, {\bf X}^3_3$ and ${\bf X}^3_4$ are obtained as
\begin{eqnarray}
& & I^3_1 = -\frac{B}{3 A \Phi} \left( \frac{\dot{B}}{B} + \frac{\dot{\Phi}}{\Phi} \right), \label{b3ks-I31} \\& & I^3_2 = -\frac{2 A B \Phi}{3 \nu_1^2} \Big{[} \nu_1 \left( 2 \frac{\dot{A}}{A} + \frac{\dot{B}}{B} \right) + \nu_2 \frac{\dot{\Phi}}{\Phi} \Big{]}, \label{b3ks-I32} \\& & I^3_3 = \frac{1}{3 q \nu_1^2} \Big{[} \nu_3 \frac{\dot{A}}{A} + \nu_4 \frac{\dot{B}}{B} + \nu_5 \frac{\dot{\Phi}}{\Phi} \Big{]}, \label{b3ks-I33} \\& &  I^3_4 = - g(t) E_L, \label{b3ks-I34}
\end{eqnarray}
where the coefficients $\nu_1, \ldots, \nu_5$ are defined as
\begin{eqnarray}
& & \nu_1= 6 \lambda (B \Phi)^2 + q, \quad \nu_2= 45 \lambda (B \Phi)^2 + \frac{3 q}{2}, \label{nu12} \\
& & \nu_3= - 6 \lambda^2 (B \Phi)^4 - 4 \lambda q (B \Phi)^2 - \frac{q^2}{2}, \,\,  \qquad \label{nu3} \\& & \nu_4=  - 6 \lambda^2 (B \Phi)^4 - 3 \lambda q (B \Phi)^2 - q^2, \label{nu4} \\& &  \nu_5= \frac{9 \lambda q }{2} (B \Phi)^2 + \frac{q^2}{4}. \label{nu5}
\end{eqnarray}

\section{Exact Solutions}
\label{fieldeqns}

In order to find new exact solutions of the field equations in the cases of the above section, the algebra of the NGS generators has to be closed, and the first integrals need to be utilized. Now, we proceed in this manner.

For case (i), in the view of relation (\ref{f-degenerate}), the field equations (\ref{feq1})-(\ref{feq4}) of BI ($q=0$), BIII ($q=-1$) and KS ($q=1$) spacetimes reduce to
\begin{eqnarray}
& & \qquad \frac{\dot{B}^2}{B^2} + 3 \frac{\dot{\Phi}^2}{\Phi^2} +
2\frac{\dot{A} \dot{B}}{A B} + 2\frac{\dot{A} \dot{\Phi}}{A \Phi}+
4\frac{\dot{B} \dot{\Phi}}{B \Phi} + \frac{q}{B^2} + 6 \frac{U(\Phi)}{\Phi^2} =0, \label{feqi-1}\\& & \qquad 2 \frac{\ddot{B}}{B} + 2 \frac{\ddot{\Phi}}{\Phi} + \frac{\dot{B}^2}{B^2}
- \frac{\dot{\Phi}^2}{\Phi^2} + 4\frac{\dot{B} \dot{\Phi}}{B \Phi} + \frac{q}{B^2}
+ 6 \frac{U(\Phi)}{\Phi^2} = 0,\label{feqi-2} \\
& & \qquad  \frac{\ddot{A}}{A} + \frac{\ddot{B}}{B} + 2 \frac{\ddot{\Phi}}{\Phi} -
\frac{\dot{\Phi}^2}{\Phi^2}+ \frac{\dot{A} \dot{B}}{A B}  + 2
\left( \frac{\dot{A}}{A} + \frac{\dot{B}}{B}\right)
\frac{\dot{\Phi}}{\Phi} + 6 \frac{U(\Phi)}{\Phi^2} = 0, \label{feqi-3}
\\& & \frac{\ddot{A}}{A}+ 2\frac{\ddot{B}}{B} + 3 \frac{\ddot{\Phi}}{\Phi} +
\frac{\dot{B}^2}{B^2} + 2\frac{\dot{A}\dot{B}}{A B} + 3 \left( \frac{\dot{A}}{A} + 2\frac{\dot{B}}{B}\right)
\frac{\dot{\Phi}}{\Phi}  + \frac{q}{B^2} + 3 \frac{U'(\Phi)}{\Phi} = 0.\qquad  \,\,
\label{feqi-4}
\end{eqnarray}
We consider only case (i) here, because unfortunately in case (ii) it has proven difficult, if sometimes impossible, to find any solutions of the field equations satisfying NGSs.

\subsection{Bianchi I Spacetime}
\label{sol-b1-spacetime}

For this spacetime, we find from the above field equations that $\lambda$ vanishes in cases (i.2), (i.4), (ii.a); and the scalar field $\Phi$ is constant in case (ii.b.1). Therefore, we drop these cases and only consider cases (i.1) and (i.3).

{\bf Case (i.1)}: To get a closed algebra of the NGS generators ${\bf X}_6, {\bf X}_7$ and ${\bf X}_8$ in this case of vanishing potential, the commutator relations (\ref{commut-i1}) of Lie brackets require that
\begin{equation}
B \Phi = k, \qquad A= \left( k \ell \right)^{-3/2} B, \label{casei1-s1}
\end{equation}
where $k$ and $\ell$ are non-zero constants.
Note that we have not explicitly obtained the scalar field $\Phi$, but the metric functions $A$ and $B$ are stated in terms of the field.

The non-zero Lie brackets  of ${\bf X}_6, {\bf X}_7$ and ${\bf X}_8$ become
\begin{eqnarray}
& & [{\bf X}_6, {\bf X}_7] = \frac{1}{k}({\bf X}_2 - 3 {\bf X}_3) + \ln{(k)} \,{\bf X}_7, \,\, [{\bf X}_6, {\bf X}_8] = 3 \ell ({\bf X}_2 - {\bf X}_3) + \ln (\ell)  {\bf X}_8, \quad \nonumber
\end{eqnarray}
which give indeed a closed algebra. Then, using the fact that $E_L = 0$ because of (\ref{feq1}), eqs. (\ref{fint12-b1-i1})-(\ref{fint8-b1-i1}) yield that all the first integrals $I_1,\ldots, I_8$ vanish. Eq. (\ref{casei1-s1}) gives also
\begin{eqnarray}
& & \frac{\dot{A}}{A} = \frac{\dot{B}}{B} = - \frac{\dot{\Phi}}{\Phi}.
\end{eqnarray}
Putting these relations in the field equations (\ref{feq1})-(\ref{feq4}) we find that the field equations are identically satisfied.

{\bf Case (i.3)}: Recall that for this case the potential is $U(\Phi) = \lambda \Phi^4$, and the coupling function $F(\Phi)$ is given by (\ref{f-degenerate}). For simplicity of writing, we will rename the constants of motion found in (\ref{fint-b1-i-3}) as $I^3_1 \equiv J_1$, $I^3_2 \equiv J_2$, $I^3_3 \equiv J_3$ and $I^3_4 \equiv J_4$. The last one vanishes by (\ref{feqi-1}) and for the others we get
\begin{eqnarray}
& & \frac{\dot{A}}{A} - \frac{\dot{B}}{B} = \frac{3 J_1}{A B^2 \Phi^2}, \label{sln-i-3-1} \\& & \frac{\dot{B}}{B} + \frac{\dot{\Phi}}{\Phi} = - 3 J_2 \frac{A \Phi}{B}, \label{sln-i-3-2} \\& & 2  \frac{\dot{A}}{A} + \frac{\dot{B}}{B} + 3 \frac{\dot{\Phi}}{\Phi} = - 9 J_3 \frac{B \Phi}{A}, \label{sln-i-3-3}
\end{eqnarray}
which are actually the constraints that have to be satisfied by the field equations (\ref{feqi-1})-(\ref{feqi-4}) with $q=0$. Substitution of (\ref{sln-i-3-2}) into the field equations (\ref{feqi-3}) and (\ref{feqi-4}) gives
\begin{equation}
\frac{\ddot{A}}{A} + \frac{\ddot{\Phi}}{\Phi} -
\frac{\dot{\Phi}^2}{\Phi^2} + \frac{\dot{A} \dot{\Phi}}{A \Phi} - 6 J_2 \frac{A \Phi}{B} \left( \frac{\dot{A}}{A} + \frac{\dot{\Phi}}{\Phi} \right)  + 6 \lambda \Phi^2 = 0, \label{feqi-3-1}
\end{equation}
\begin{equation}
\frac{\ddot{A}}{A}+ \frac{\ddot{\Phi}}{\Phi} -
\frac{\dot{\Phi}^2}{\Phi^2} + \frac{\dot{A}\dot{\Phi}}{A \Phi} - 12 J_2 \frac{A \Phi}{B} \left( \frac{\dot{A}}{A} + \frac{\dot{\Phi}}{\Phi}\right)
+ 9 (J_2)^2 \left( \frac{A \Phi}{B} \right)^2 + 12 \lambda \Phi^2 = 0. \label{feqi-4-1}
\end{equation}
Then, subtracting Eq. (\ref{feqi-3-1}) from Eq. (\ref{feqi-4-1})), one gets
\begin{equation}
-6 J_2 \frac{A \Phi}{B} \left( \frac{\dot{A}}{A} + \frac{\dot{\Phi}}{\Phi} \right) + 9 (J_2)^2 \left( \frac{A \Phi}{B} \right)^2 + 6 \lambda \Phi^2 = 0,
\end{equation}
which yields the constraint equation
\begin{equation}
\left( A \Phi \right)^{\bf .} = 3 J_2 \frac{A^2 \Phi^2}{2 B} + \frac{\lambda}{J_2} B \Phi^2, \label{sln-i-3-6}
\end{equation}
where $J_2 \neq 0$. Here we need to correct a mistake in our previous study \cite{camci}, where we have used $c_0$ instead of the constant of motion $J_2$; and they are related by $c_0 = -3 J_2$. Also, it can be seen that the Eq. (42) of that reference is erroneously calculated, the correct one is  Eq.(\ref{feqi-4-1}). Thus, the Eqs. (43) and (47) of Ref. \cite{camci} do not apply and the solution (49) does not exist.

Using the Eqs. (\ref{sln-i-3-1}) and (\ref{sln-i-3-2}) in (\ref{sln-i-3-3}) we find another constraint relation
\begin{equation}
2 J_1 - 3 J_2 A^2 B \Phi^3 + 3 J_3 (B \Phi)^3 = 0. \label{sln-i-3-4}
\end{equation}
Considering (\ref{sln-i-3-1})-(\ref{sln-i-3-3}) in the field equations (\ref{feqi-1}) and (\ref{feqi-2}) gives the third constraint relation
\begin{equation}
- 6 J_1 I_2 + 9 J_2 A^2 B \Phi^3 + 2 \lambda (B \Phi)^3 = 0. \label{sln-i-3-5}
\end{equation}
The remaining field equations (\ref{feqi-3}) and (\ref{feqi-4}) are identically satisfied using the obtained relations (\ref{sln-i-3-4}) and (\ref{sln-i-3-5}). Now, we use the transformation of the time coordinate by $dt = (B/A \Phi) d\tau$ in the above Eqs. (\ref{sln-i-3-2}) and (\ref{sln-i-3-6}). Further, after integration with respect to the new time coordinate we find
\begin{eqnarray}
& & A^2 = \frac{2 }{ 9 a (J_2)^3 \Phi^2} \left( 3 J_1 J_2 e^{ 3 J_2 \tau } - \lambda a^3 e^{- 6 J_2 \tau} \right), \label{sln-a} \qquad \\& & B  = a \, \frac{e^{ -3 J_2 \tau}}{\Phi}, \label{sln-b}
\end{eqnarray}
where $a$ is a constant of integration. Again, we have given the metric functions in terms of the scalar field.

\subsection{Bianchi III and Kantowski-Sachs Spacetimes}
\label{sol-b3ks-spacetime}

We only show treatments of the cases (i.1) and (i.3) in this section.

{\bf Case (i.1)}: Using the form $F= \epsilon \Phi^2 /12$ of the coupling function, and the vanishing of the potential $U(\Phi)$, the first integrals of this subcase give
\begin{eqnarray}
& & \frac{\dot{A}}{A} = - \frac{\dot{\Phi}}{\Phi} + \frac{3 I_3}{A B^2 \Phi^2}, \label{b13ks-i1-1} \\& &  \frac{\dot{B}}{B} = - \frac{\dot{\Phi}}{\Phi} + \frac{3 I_7}{A B \Phi}- \frac{6 I_3}{A B^2 \Phi^2}, \label{b13ks-i1-2}
\end{eqnarray}
where $I_3$ and $I_7$ are constants of motion. Substituting the above relations into the field equations (\ref{feq1})-(\ref{feq4}) for BIII ($q=-1$) and KS ($q=1$) spacetimes, we find the following constraint equation
\begin{eqnarray}
& &  q A^2 B \Phi^3 + 9 (I_7)^2 B \Phi - 18 I_3 I_7 =0. \label{ceq-i1-1}
\end{eqnarray}
In order to find solution of the Eqs. (\ref{b13ks-i1-1}) and (\ref{b13ks-i1-2}), we use the transformation of time coordinate by $dt = A B^2 \Phi^2 d\tau$, and find
\begin{eqnarray}
& & A = A_0 \frac{\exp(3 I_3 \tau)}{\Phi}, \quad B = \frac{ 2 I_3}{[I_7 + 2 B_0 \exp(6 I_3 \tau)] \Phi},\qquad \quad \label{sln-b3-s1}
\end{eqnarray}
where $A_0$ and $B_0$ are integration constants. Putting these into the constraint equation (\ref{ceq-i1-1}) yields
\begin{equation}
I_7 = \frac{q A_0^2}{18 B_0},
\end{equation}
with $B_0 \neq 0$.

{\bf Case (i.3)}: In this case, we use the first integral (\ref{b3ks-I31}) which gives
\begin{equation}
\frac{\dot{B}}{B} + \frac{\dot{\Phi}}{\Phi} = - \frac{3 J_1 A \Phi }{B}, \label{sol-i3}
\end{equation}
where we have renamed $I^3_1 \equiv J_1$, as in the case (i.3) of subsection \ref{sol-b1-spacetime}. Similar to the discussion there, using (\ref{sol-i3}) in the field equations (\ref{feqi-3}) and (\ref{feqi-4}) gives
\begin{equation}
\frac{\ddot{A}}{A} + \frac{\ddot{\Phi}}{\Phi} -
\frac{\dot{\Phi}^2}{\Phi^2} + \frac{\dot{A} \dot{\Phi}}{A \Phi} - 6 J_1 \frac{A \Phi}{B} \left( \frac{\dot{A}}{A} + \frac{\dot{\Phi}}{\Phi} \right)  + 6 \lambda \Phi^2 = 0, \label{feq-i3-1}
\end{equation}
\begin{equation}
\frac{\ddot{A}}{A}+ \frac{\ddot{\Phi}}{\Phi} -
\frac{\dot{\Phi}^2}{\Phi^2} + \frac{\dot{A}\dot{\Phi}}{A \Phi} - 12 J_1 \frac{A \Phi}{B} \left( \frac{\dot{A}}{A} + \frac{\dot{\Phi}}{\Phi}\right)  + 12 \lambda \Phi^2 +  9 (J_1)^2 \left( \frac{A \Phi}{B} \right)^2 + \frac{q}{B^2} = 0.
\label{feq-i3-2}
\end{equation}
Subtracting these equations one gets
\begin{equation}
-6 J_1 \frac{A \Phi}{B} \left( \frac{\dot{A}}{A} + \frac{\dot{\Phi}}{\Phi} \right) + 9 (J_1)^2 \left( \frac{A \Phi}{B} \right)^2 + 6 \lambda \Phi^2 + \frac{q}{B^2}= 0,
\end{equation}
which can be brought into the form
\begin{equation}
\left( A \Phi \right)^{\bf .} = 3 J_1 \frac{(A \Phi)^2}{2 B} + \frac{\lambda}{J_1} B \Phi^2 +  \frac{q}{6 J_1 B}, \label{sln-i-3-6-b3}
\end{equation}
where $J_1 \neq 0$. Then, we consider the remaining first integrals (\ref{b3ks-I32}) and (\ref{b3ks-I33}) to obtain
\begin{eqnarray}
& & 2 \nu_1 \frac{\dot{A}}{A} + (\nu_2 - \nu_1) \frac{\dot{\Phi}}{\Phi} = \frac{3 \nu_1}{B} \left( J_1 A \Phi + \frac{J_2 \nu_1}{2 A \Phi}\right), \qquad \label{feq-i3-a}\\& & \nu_3 \frac{\dot{A}}{A} + (\nu_5 - \nu_4) \frac{\dot{\Phi}}{\Phi} = 3 J_1 \nu_4  \frac{A \Phi}{B} +3 q J_3 \nu_1^2, \label{feq-i3-b}
\end{eqnarray}
where $J_2$ and $J_3$ stand for  $I^3_2$ and $I^3_3$, respectively, as above. Recall that $\nu_i \, (i=1,\ldots,5)$ are functions of the combination $(B\Phi)$, given in (\ref{nu12})-(\ref{nu5}).

Using the time transformation $dt = (B/A\Phi) d\tau$ in (\ref{sol-i3}) and (\ref{sln-i-3-6-b3}), we find
\begin{eqnarray}
& &  (A \Phi)^2 = - \frac{1}{c_0^2} \left[ q + 2 \lambda a^2 \exp(2 c_0 \tau) \right] + b \exp(-c_0 \tau), \,\, \qquad \label{sln-b3-s2-1} \\& & B \Phi = a \exp( c_0 \tau), \label{sln-b3-s2-2}
\end{eqnarray}
where $c_0 = - 3 J_1$ (see Eq.(46) of ref.\cite{camci}), $a$ and $b$ are constants of integration. Note that both combinations $(A\Phi)$ and $(B\Phi)$  are now functions of $\tau$.

Under this time transformation,  the Eqs. (\ref{feq-i3-a}) and (\ref{feq-i3-b}) become
\begin{eqnarray}
& & 2 \nu_1 \frac{A'}{A} + (\nu_2 - \nu_1) \frac{\Phi'}{\Phi} = 3 \nu_1 \left( J_1  + \frac{J_2 \nu_1}{2 (A \Phi)^2}\right), \qquad \label{feq-i3-a2}\\& & \nu_3 \frac{A'}{A} + (\nu_5 - \nu_4) \frac{\Phi'}{\Phi} = 3 J_1 \nu_4  + 3 q J_3 \nu_1^2 \frac{B}{A \Phi}. \label{feq-i3-b2}
\end{eqnarray}
where the prime $(')$ denotes differentiation with respect to $\tau$.
Eliminating the term $A' / A$ from these equations, it follows that
\begin{equation}
f_1 (\tau) \Phi' + f_2 (\tau) \Phi = f_3 (\tau), \label{phi-i3}
\end{equation}
where $f_1 (\tau), f_2 (\tau)$ and $f_3 (\tau)$ are given by
\begin{eqnarray}
& & f_1 (\tau) = \frac{\nu_3}{2 \nu_1}(\nu_1 - \nu_2)  - \nu_4 + \nu_5, \\& & f_2 (\tau) = \frac{3 \nu_3}{2} \left[ J_1 + \frac{J_2 \nu_1}{2 (A\Phi)^2}\right] - 3 J_1 \nu_4, \\& & f_3 (\tau) = 3 a q J_1 \nu_1^2 \frac{\exp(c_0 \tau)}{A \Phi}.
\end{eqnarray}
The equation (\ref{phi-i3}) is a linear ordinary differential equation of first order and it has a solution of the form
\begin{equation}
\Phi (\tau) = \mu (\tau)^{-1} \left[ \int{f_3 (\tau) \mu (\tau) d\tau} + \Phi_0\right],  \label{phi-i3-s1}
\end{equation}
where $\mu (\tau) = \exp\left[ \int{ d\tau f_2 (\tau) / f_1 (\tau)}\right]$. We have not been able to solve the above integrals yet.

\section{Concluding remarks}
\label{conc}

In the present work, we have used the Noether gauge symmetry approach to search the Noether symmetries of Lagrangian (\ref{lag-b13ks}) for BI, BIII and KS spacetimes in two theories involving scalar fields non-minimally coupled to gravity, the $\Phi^{2}R$ theory motivated by the induced theory of gravity, and the more standard $F(\Phi) = 1- \zeta \Phi^2$ theory. A particular form of the former, $F(\Phi)=\frac{\epsilon}{12}\Phi^{2}$, arises in the case of the vanishing of the Hessian determinant of the Lagrangian, a.k.a. the degenerate case. We have shown that a number of Noether gauge symmetries for BI, BIII, and KS spacetimes exist and each of them are related to a constant of motion. Using the two coupling functions $F(\Phi)$ and the choices given in Tables \ref{b1-t1} or \ref{b3ks-t1} for the potential $U(\Phi)$, we used the first integrals obtained through the NGSs to solve the field equations for BI, BIII and KS spacetimes. In some cases, the NGSs and the first integrals are explicitly elaborated in the paper, for other cases, the NGS vector fields and corresponding first integrals are collected in Tables \ref{b1-t2} and \ref{b3ks-t2}. The maximum number of NGS generators is found to vary from two to eight in the cases considered.  Here we also correct a mistake in our previous study \cite{camci}, namely we find that NGSs for BI, BIII and KS spacetimes {\em do} exist, contrary to the claim of that work.

In both of the cases (i) and (ii), it is possible to find particular exact solutions for the system of field equations (\ref{feq1})-(\ref{feq4}), obtaining the explicit behaviour of the scale factors $A(t)$ and $B(t)$. For BI spacetime, we have found the exact solutions (\ref{casei1-s1}) in case (i.1) and (\ref{sln-a})-(\ref{sln-b}) in case (i.3). Also, for BIII and KS spacetimes, we have found the exact solutions (\ref{sln-b3-s1}) in case (i.1) and (\ref{sln-b3-s2-1})-(\ref{sln-b3-s2-2}) in case (i.3). As it is clearly seen in the solutions (\ref{casei1-s1}), (\ref{sln-a})-(\ref{sln-b}), (\ref{sln-b3-s1}) and (\ref{sln-b3-s2-1})-(\ref{sln-b3-s2-2}) in terms of the conformal time $\tau$, both of the scale factors $A(\tau)$ and $B(\tau)$ are proportional to the inverse of the scalar field $\Phi$. We do not have the explicit form of the scalar field $\Phi$ in the cases (i.1) and (i.3) for BI spacetime, and in the case (i.1) for BIII and KS spacetimes. We have found the scalar field in terms of a quadrature, eq. (\ref{phi-i3-s1}), in case (i.3) for BIII and KS spacetimes, but we have been unable to solve the integral.

In cosmology, where observations show that the universe is both spatially homogeneous and isotropic to a high degree, these models are of physical interest because of their spatial homogeneity. With regard to the remaining anisotropies in the Cosmic Microwave Background (CMB) radiation, the Bianchi type spacetimes can serve as models more realistic than FLRW. Therefore the interest in these models has increased in the context of the analysis and interpretation of the WMAP data \cite{wmap}.

Even though the observed universe seems to be almost isotropic on large scales, the early universe could be anisotropic \cite{ellis}. Quantum gravitational aspects may be observed in the anisotropy spectrum of CMB, and future observations may be able to detect the contribution of \emph{inflation} generated in the early universe \cite{capo2011}. Thus the models of inflationary cosmology could be Bianchi models. It is also interesting to note that although the inflationary scenarios originally proposed using minimally coupled scalar fields, some later models have widely used non-minimally coupled scalar fields $\Phi$ in relation with specific inflationary scenarios \cite{barroso, hwang}; and in the present study, the new exact solutions of BI, BIII and KS spacetimes have just such an interesting property, namely, the scale factors include a non-minimally coupled scalar field $\Phi$.

\section*{Acknowledgements}

We would like to thank Dr. \.{I}. Semiz for his detailed reading, useful comments and suggestions on this manuscript. This work was supported by The Scientific Research Projects Coordination Unit of Akdeniz University (BAP). Project Number: 2013.01.115.003.

\section*{References}


\end{document}